\documentclass[]{spie}  %>>> use for US letter paper
\usepackage{amsmath,amsfonts,amssymb}
\usepackage{graphicx}
\usepackage[colorlinks=true, allcolors=blue]{hyperref}

\title{Detector systems engineering for extremely large instruments}

\author[a*]{Elizabeth M. George}
\author[a]{Naidu Bezawada}
\author[a]{Derek Ives}
\author[a]{Leander Mehrgan}
\author[a]{Matteo Accardo}
\author[a]{Domingo Alvarez}
\author[a]{Martin Brinkmann}
\author[a]{Ralf Conzelmann}
\author[a]{Claudio Cumani}
\author[a]{Mark Downing}
\author[a]{Max Engelhardt}
\author[a]{Marcus Haug}
\author[a]{Joshua Hopgood}
\author[a]{Christoph Geimer}
\author[a]{Olaf Iwert}
\author[a]{Barbara Klein}
\author[a]{Christopher Mandla}
\author[a]{Eric M\"{u}ller}
\author[a]{Suzanne Ramsay}
\author[a]{Javier Reyes}
\author[a]{Mathias Richerzhagen}
\author[a]{Beno\^{i}t Serra}
\author[a]{Matthias Seidel}
\author[a]{J\"{o}rg Stegmeier}
\author[a]{Mirko Todorovic}
\affil[a]{European Southern Observatory, Karl-Schwarzschild-Str. 2, Garching 85748, GERMANY}

\authorinfo{Corresponding author: Elizabeth M. George, egeorge@eso.org}

\begin{document} 
\maketitle

\begin{abstract}
The scientific detector systems for the ESO ELT first-light instruments, HARMONI, MICADO, and METIS, together will require 27 science detectors: seventeen 2.5 $\mu$m cutoff H4RG-15 detectors, four 4K x 4K 231-84 CCDs, five 5.3 $\mu$m cutoff H2RG detectors, and one 13.5 $\mu$m cutoff GEOSNAP detector. This challenging program of scientific detector system development covers everything from designing and producing state-of-the-art detector control and readout electronics, to developing new detector characterization techniques in the lab, to performance modeling and final system verification. We report briefly on the current design of these detector systems and developments underway to meet the challenging scientific performance goals of the ELT instruments. 
\end{abstract}


\section{INTRODUCTION}
\label{sec:intro}  % \label{} allows reference to this section

The Extremely Large Telescope (ELT) is now in the construction phase, along with the first set of instruments, HARMONI, MICADO, and METIS.\cite{ramsay20, ramsay21} These instruments cover a wide range of science cases designed to exploit the full power of the 39-m ELT to collect photons from extremely faint objects and resolve systems with extremely small spatial scales.  ESO is responsible for delivering the scientific detector systems of the ELT first light instruments to the instrument consortia, which together will require 27 science detectors: seventeen 2.5 $\mu$m cutoff H4RG-15 detectors, four 4K x 4K 231-84 CCDs, five 5.3 $\mu$m cutoff H2RG detectors, and one 13.5 $\mu$m cutoff GEOSNAP detector. All of these instruments and the ELT itself additionally include AO systems and the wave front sensing detectors that go with them, which are covered elsewhere in these proceedings.\cite{marchetti20, reyes20}

This challenging program of scientific detector system development and delivery covers everything from designing and producing state-of-the-art detector control and readout electronics, to developing new detector characterization techniques in the lab, to performance modeling and final system verification. In section \ref{sec:InsFP} we describe three instruments and their focal planes. In sections \ref{sec:coldelectonics} and \ref{sec:NGCII} we report briefly on the current design of the cold detector electronics and the development of the next generation of ESO's detector controller, the NGCII. The test program is described in section \ref{sec:test} and our collaborative work on the detector simulation framework, Pyxel,\cite{lucsanyi18, prodhomme20} is described in section \ref{sec:modeling}.

\section{ELT first light instrument focal planes}
\label{sec:InsFP}

\subsection{HARMONI}
\label{sec:HARMONI}
\begin{figure} [htbp]
   \begin{center}
   \begin{tabular}{c}
     \includegraphics[width=1.0\textwidth]{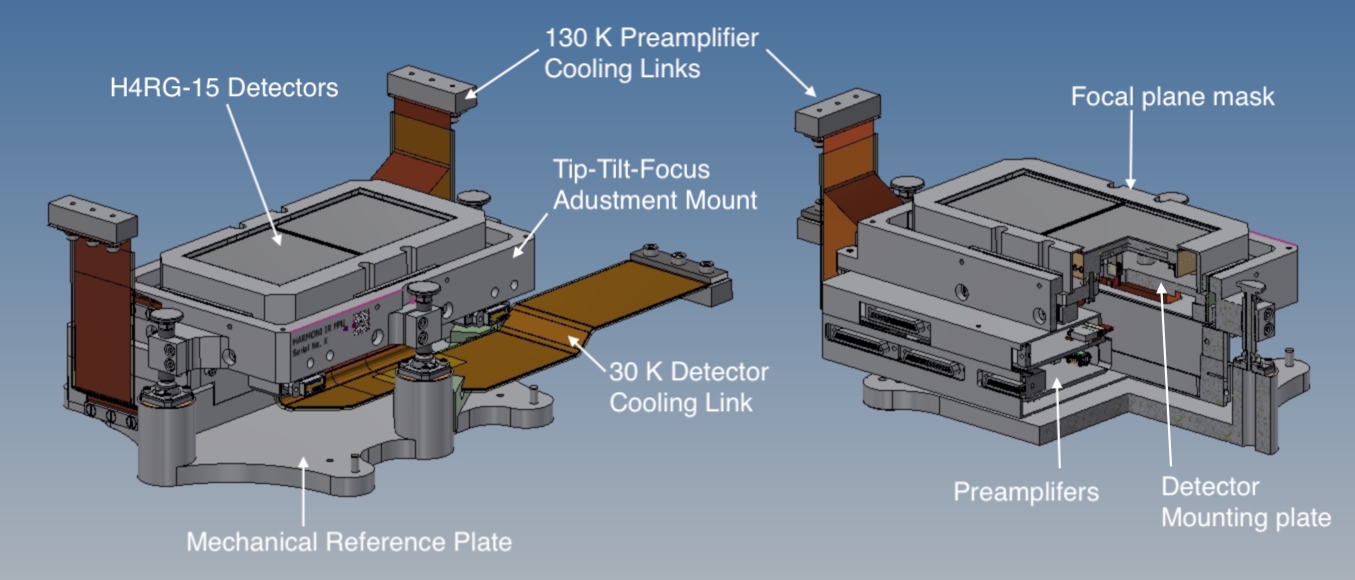}
   \end{tabular}
   \end{center}
   \caption[HARMONI IR Focal Plane CAD model] 
   { \label{fig:HARMONI_FP}  \centering
	CAD model of one of four HARMONI IR focal planes. The two visible focal planes are similar in concept but have a different design due to the differing connections required by the detectors. The detectors are precisely co-aligned on the detector mounting plate to match the slit gap in the spectrograph output field. The position of the detectors is referenced to the Mechanical Interface plate to allow spectrograph alignment.}
   \end{figure}

The HARMONI\cite{thatte20} instrument is an integral field spectrograph that covers the wavelength range 0.47-2.45 $\mu$m using two different detector types. There are four spectrographs in HARMONI, four of which have 2x1 mosaics of IR detectors, and two of which additionally have 2x1 mosaics of visible detectors. Figure \ref{fig:HARMONI_FP} shows a CAD model of a single HARMONI IR focal plane with various parts labeled. Eight Teledyne H4RG-15 detectors cover the astronomical I,Y,J,H, and K bands (0.81-2.45 $\mu$m) with several configurations providing different wavelength ranges and spectral resolutions (R $\sim$ 3,500-17,000).  Four Te2v CCD 231-84 detectors cover half of the field of view of the IR detectors the visible bands (0.47-0.83 $\mu$m) at a fixed resolution of $\sim$ 3,500. 

HARMONI will serve a wide variety of science cases, from kinematics of faint galaxies to the dynamics of circumstellar disks. This requires low noise and dark current and high stability for long exposures. Due to the fact that there are 11 different configurations of spectra that will be imaged onto the IR focal planes, persistence of bright sky lines between exposures is of particular concern, as the sky lines will appear in different parts of the detector in every configuration. In addition, the detectors should have excellent cosmetic performance as any cosmetic defect would remove information at a fixed wavelength in a given spaxel for a given configuration. The detectors need to be aligned precisely in height and in translation and rotation to ensure that the full spectra for all configurations is imaged onto the detector active areas. Initial testing of the H4RG-15 detectors for the MOONS project\cite{ives20} indicates that performance similar to that obtained previously with H2RG detectors is achievable and they will meet HARMONI's specifications.

\subsection{MICADO}
\label{sec:MICADO}
\begin{figure} [htbp]
   \begin{center}
   \begin{tabular}{c}
     \includegraphics[width=1.0\textwidth]{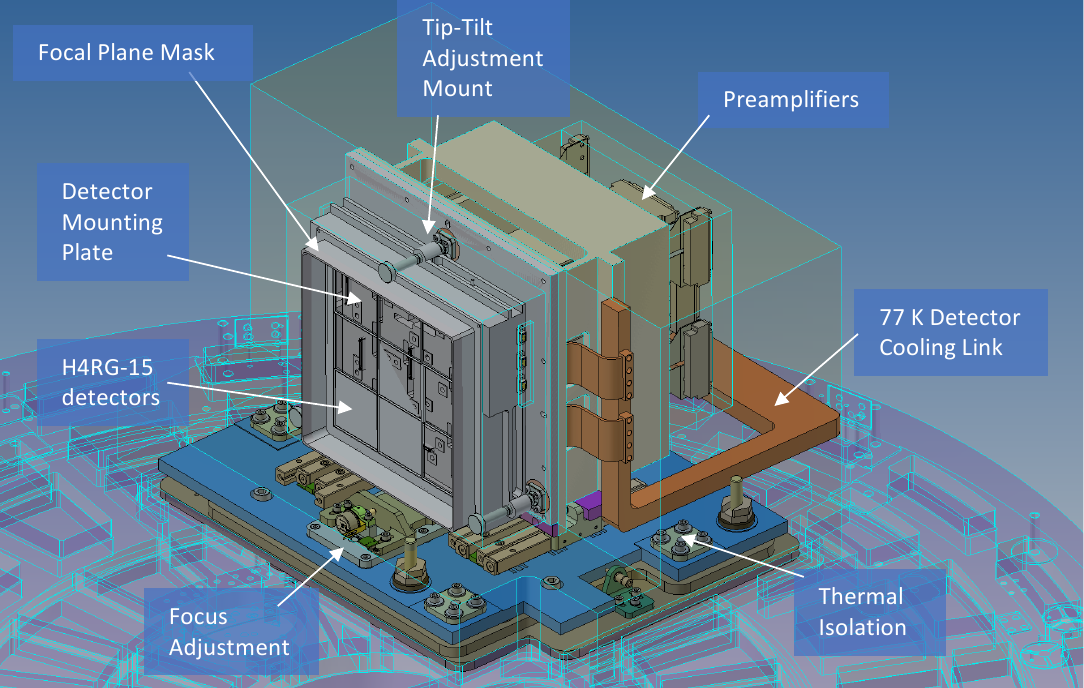}
   \end{tabular}
   \end{center}
   \caption[HARMONI IR Focal Plane CAD model] 
   { \label{fig:MICADO_FP}  \centering
	CAD model of the MICADO 3 x 3 mosaic focal plane.  The detectors are precisely co-aligned on the detector mounting plate to provide a flat focal plane for the camera. All detector cables are routed out the back of the focal plane.}
   \end{figure}

MICADO\cite{davies18} is a near-infrared imager and low-resolution spectrograph covering the wavelength range of $\sim$ 0.8 - 2.45 $\mu$m, that will work together with the AO system MAORY to achieve diffraction-limited performance.\cite{ciliegi18} The focal plane consists of nine H4RG-15 detectors in a 3 x 3 mosaic to cover a field of 53' x 53' in the larger of the two plate scales (1.5 and 4 mas/px). Figure \ref{fig:MICADO_FP} shows a CAD model of the MICADO focal plane with various components labeled.

While MICADO covers a range of science cases as a diffraction-limited camera on the largest of the ELTs, a primary goal of the instrument is the high-precision astrometric mode. With a requirement of 50 $\mu$as accuracy and a goal of 10 $\mu$as, we must be able to determine the position of stars on the detector to within 1/150th of a pixel. This requires not only outstanding stability of the detectors, but characterization of any detector effects such as Inter Pixel Capacitance (IPC) or persistence that may apparently result in a measured shift in charge position. As is the case for HARMONI, the initial testing of the MOONS H4RG-15 detectors at ESO is already providing insight into the properties of the H4RG-15 detectors that could impact MICADO (see Ives et al., these proceedings\cite{ives20}).

\subsection{METIS}
\label{sec:METIS}
\begin{figure} [htbp]
   \begin{center}
   \begin{tabular}{c}
     \includegraphics[width=1.0\textwidth]{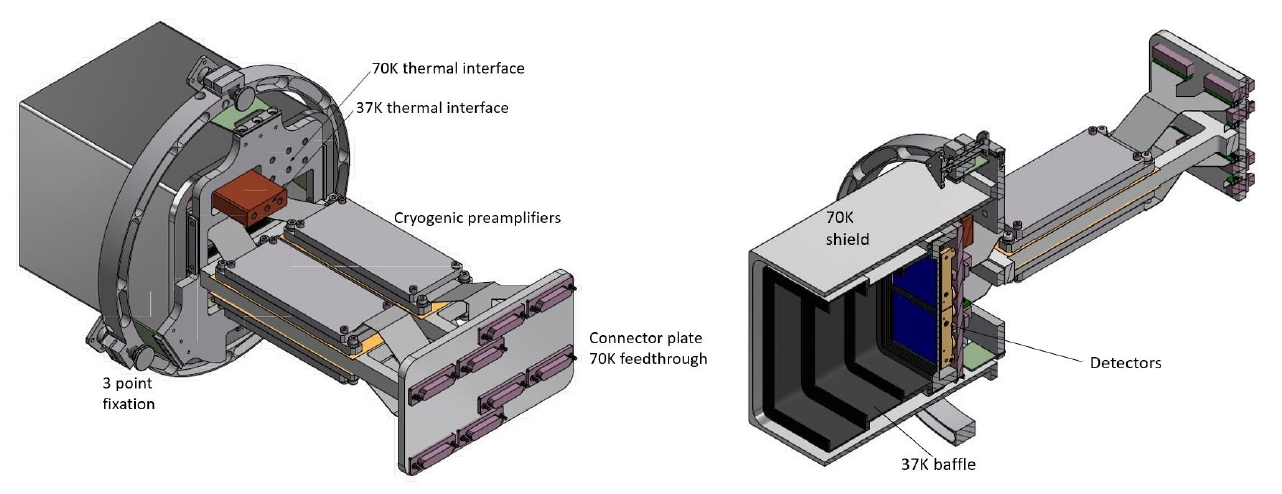}
   \end{tabular}
   \end{center}
   \caption[HARMONI IR Focal Plane CAD model] 
   { \label{fig:METIS_FP}  \centering
	CAD model of the METIS 2 x 2 mosaic LMS (IFU) focal plane. The two imager focal planes are similar in concept but only have one detector each. The design is based on the very similar H2RG mosaic used in the HAWK-I instrument.}
   \end{figure}

METIS is an instrument operating in the thermal infrared covering the wavelength range from 3-13.5 $\mu$m using two different modules. A mosaic of four 5.3 $\mu$m cutoff H2RG detectors form the focal plane for the high resolution (R $\sim$ 100,000) IFU-fed spectrograph, which will cover the L and M bands (LMS). A single 5.3 $\mu$m cutoff H2RG detector covering the L and M bands along with a 2k x 2k Teledyne GEOSNAP detector covering the N band are used for the diffraction-limited imager (in two separate focal planes). This module includes low/medium resolution spectroscopy and coronography for high-contrast imaging. Figure \ref{fig:METIS_FP} shows a mock-up of the METIS LMS focal plane.

This combination of detectors allows the METIS instrument to perform its primary science case of exoplanet detection and characterization. High resolution spectroscopy requires the lowest noise performance, high stability, and a strong understanding of detector characteristics such as persistence or IPC that can subtly impact recorded spectra. The detectors must also be precisely co-aligned on the focal plane. The H2RG detectors meet all these requirements and have a long heritage in many instruments at ESO and world-wide. %(anything else to add? cite some papers? Any hard numbers?) 

Imaging in the thermal infrared requires a high detector operation speed, a large full well capacity, and low 1/f noise. In the LM bands, the H2RG detector meets these requirements while operating in fast mode at 5 Mpix/s readout rate, while also allowing for low-noise operation for long-slit spectroscopy. In the N band, the GEOSNAP detector was chosen. While the GEOSNAP detector has higher dark current than a comparable Si:As detector would, the lack of Excess Low Frequency Noise (ELFN) and higher QE means that when operated with the METIS internal chopper to suppress 1/f noise, it provides the highest S/N in the shortest exposure time. The higher operating temperature of GEOSNAP also simplifies cryostat design. Initial testing of the GEOSNAP detector within the METIS consortium shows promising performance.%(cite papers here, or maybe add hard numbers? Anything else to add?)

\subsection{Future ELT instruments}
\label{sec:futureins}
While not the subject of this paper, we should not lose sight of the fact that there are three more instruments in the ELT instrumentation plan: HIRES, MOSAIC, and ELT-PCS.\cite{ramsay20} HIRES\cite{marconi20} is a high resolution spectrograph (R $\sim$100,000), and MOSAIC\cite{jagourel18, hammer20} is a multi-object spectrograph with resolving power of R $\sim$ 5,000 and 20,000). The conceptual designs for both instruments are complete, and both will require large-format visible and near infrared detectors, covering wavelengths ranging from $\sim$0.37 - 2.4 $\mu$m. Both instruments, but in particular HIRES require very high stability and very well understood detector systematics. 

As the designs progress into the preliminary design phase, detectors that will be available on the market or developed in the next decade will need to be evaluated against performance requirements of these instruments. We will gain valuable information through our detector work on the first 3 ELT instruments and new instruments for the VLT such as ESPRESSO and MOONS. Additionally new developments on scientific CMOS sensors and curved detectors may widen the options available for detectors on the next generation of instruments.

\section{Cryogenic mechanics}
\label{sec:cryomechanics}
Figures \ref{fig:HARMONI_FP}, \ref{fig:MICADO_FP} and \ref{fig:METIS_FP} all show the mechanical designs for the cryogenic focal planes, which are responsible for the precise alignment, stability, and temperature control of the detectors. The designs all use some form of three-point mount for tip-tilt and focus adjustment, and some instruments also include motorized focus stages. The different parts of the focal planes also operate at different temperatures, as the detectors, preamplifiers, baffles, and mechanical interfaces could have temperature differences of over 100 K. Therefore an important aspect of this design work is the thermal and mechanical simulations that are carried out as part of the design process. Thermal control is achieved for the focal planes as units with the detectors co-mounted on a single plate, and of particular concern is ensuring uniformity of operating temperature across the focal plane and the alignment of the detectors at their cryogenic temperatures based room temperature alignment.

Once the focal planes are manufactured, the detectors need to be carefully aligned to each other and to a mechanical reference point. We use several tools to perform this task at ESO, such as a mechanical coordinate measurement machine (CMM), optical microscopes, and scanning laser distance measurements to achieve the accuracies required by the instruments.

\section{Cold electronics}
\label{sec:coldelectonics}
One of the challenges of extremely large instruments is that the detector focal planes are often buried deep inside cryostats that can be the size of a typical laboratory. This means that it is necessary for detector preamplifiers to operate at cryogenic temperatures. Additionally cryogenic cabling between the focal plane and detector controller outside the cryostat is longer than in previous instruments. This requires particular care when designing the cold electronics to ensure good performance and low noise. 

In the last two years at ESO, we have performed extensive development and testing on our HxRG cryogenic preamplifiers (see Bezawada et. al, these proceedings\cite{bezawada20}). The results of these tests show that the buffered output mode of the HxRG detectors has several advantages in terms of performance including lower crosstalk and faster readout speeds over the unbuffered mode that we have used in previous ESO instruments. For the HxRG detectors in MICADO, METIS, and HARMONI we will use a single standard cryogenic preamplifier design that is optimized for buffered output operation. The MOONS H4RG preamplifier is the first buffered HxRG preamplifier that will be delivered to an ESO instrument (see Ives et al., these proceedings\cite{ives20}).

The CCDs in the HARMONI instrument are buried deep in the cryostat, and thus we are also developing a cryogenic preamplifier for the CCD231-84 based on our experience developing a cryogenic CCD preamplifier for the MOONS project. 

\section{Detector controller}
\label{sec:NGCII}
The detector systems group at ESO has a long tradition of developing detector controllers for instruments on ESO's telescope. Perhaps the best-known of these is the NGC, which is a highly flexible and configurable controller currently used to control a variety of detector numbers and types in ESO's instruments, for example, the two 9k x 9k CCDs in ESPRESSO, the twenty-four 4k x 4k CCDs in MUSE, the three H2RGs in KMOS, and the four 1k x 1k SAPHIRA detectors on GRAVITY. The power of this controller has always been its high performance, configurability, and ability to synchronize the operation of many detectors. Having a single controller type has also significantly eased operation and maintenance of the detector systems at ESO's observatories.
 
Following on the great success of NGC, the detector systems group is now developing the next generation of NGC, the NGCII. The NGCII is based on the industry-standard MicroTCA.4 that is used for beamline control in many high-energy physics experiments. This change in base architecture allows the team to reconfigure the NGC concept to be more modular, allowing for more freedom in optimization for operating different types of detectors. Importantly, this change is also necessary to conform to the interfaces and modern network infrastructure in the ELT and also allows better synchronization of multiple detectors and more precise time-stamping of detector exposures using the PTP protocol. Finally, NGCII will simplify system architecture for instruments with large numbers of detector of different types.

A new version of the detector control software (DCS) is currently under development to meet the ELT software standards and use new features such as the Data Display Tool, which allows data from one detector to be simultaneously accessed by many systems. 

\section{Test program}
\label{sec:test}
To deliver 27 scientific detectors and their control electronics in the next five years, our test program needs to be highly efficient. One benefit of designing all of the scientific detector systems for the 3 instruments simultaneously is that it has allowed us to standardize individual components such as connectors and cables as well as systems-level designs such as readout architecture, while still retaining flexibility in the designs to meet the needs of each individual instrument. For example, as seen in figures \ref{fig:HARMONI_FP} and \ref{fig:MICADO_FP}, the form factors of the H4RG-15 preamplifiers for HARMONI and MICADO are different due to the space constraints in the two instruments, however, the electrical design and connectors are identical, allowing for the testing of the various hardware in single test setups.

The heart of our test program is the Facility for Infrared Array Testing (FIAT)\cite{lizon16}. FIAT is a cryogenic test facility in which we will test all 17 H4RG-15 detectors for the ELT instruments under identical conditions. FIAT contains astronomical bandpass filters allowing for imaging from 0.8-5.3 microns, and allows for detector operation temperatures ranging from 35-100 K. As MICADO and HARMONI plan to operate at 80 and 40 K respectively, it is particularly important to measure all detector characteristics at both temperatures. In addition to full operational flexibility and engineering-level control of the detectors, the facility is designed to run with automated templates to collect standardized data sets, with the data going directly to the ESO archive, allowing for long-term monitoring of the detector characteristics from the first acceptance tests after delivery to end-of-life on the telescope. The test plan includes full characterization of standard detector effects such as dark current, noise, cosmetics, persistence, IPC, non-linearity, and PRNU. The data will also be available to the instrument consortia to use in their modeling before the instrument ever goes on sky. 

\begin{figure} [htbp]
   \begin{center}
   \begin{tabular}{c}
     \includegraphics[width=1.0\textwidth]{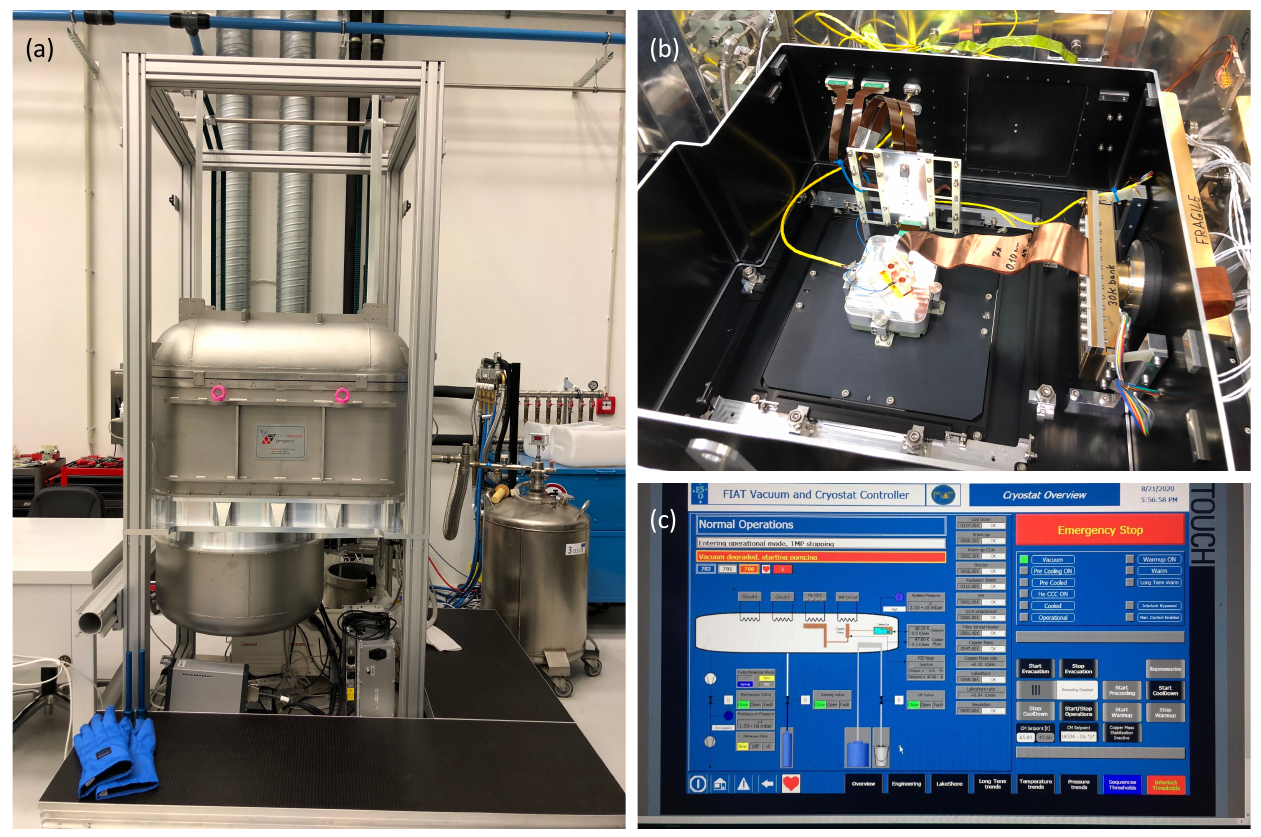}
   \end{tabular}
   \end{center}
   \caption[Photos of FIAT in final acceptance testing] 
   { \label{fig:FIAT_montage}  \centering
	Photos of FIAT in final acceptance testing. (a) FIAT cryostat in mounting frame. The cryostat cover is lifted via a weighted pulley system to ease detector integration. (b) Inside the detector chamber. The detector faces downwards and cold electronics are easily accessible. (c) The PLC control panel for automated cryostat operation.}
   \end{figure}

FIAT has achieved first light at ESO, and is currently undergoing acceptance testing before final delivery to the detector group. Figure \ref{fig:FIAT_montage} shows a few photos of FIAT in acceptance testing, with a focus on what the detector engineers will encounter when using the facility. The complex cryogenic optics and mechanics are all located in the lower part of the cryostat and will not be exposed when the cryostat is opened to exchange detectors, allowing for rapid turnaround in testing (see Lizon et al. 2016\cite{lizon16} for the optical and mechanical design).

There are several other test cryostats at ESO that will be used to test detectors for the ELT instruments, the Mosaic Test Facility (MTF) is a low-background cryostat specifically designed to test the 5.3 $\mu$m cutoff H2RG detectors used by METIS. A new visible detector test bench is currently in commissioning that will be used to test the HARMONI CCDs. Finally, we will cryogenically test all of the cold electronics and cables in another test cryostat specifically designed to do these unit tests to ensure functionality and uniform performance of the \textgreater 1000 preamplifier output channels required by the various focal planes.

\section{Modeling}
\label{sec:modeling}
As the science cases become ever more complex, an ever larger part of the data analysis and scientific interpretation relies on precise knowledge of the characteristics of the instruments, including detectors. Every ELT instrument has an instrument numerical model, a full simulation of the instrument from photons entering the instrument to detection at the detector. Additionally, advanced instrument simulations are routinely used in exposure time calculators for observation planning. 

As part of the ELT Working Groups, we are working together with colleagues at ESA on the PyXel detector simulator.\cite{prodhomme20} This open-source python-based detector simulation framework allows anyone to implement models of various detector effects and apply them to the simulation pipeline. The synthetic readouts produced by the simulation can be used as the last step in the instrument numerical models to create realistic detector readouts whilst changing the parameters of various detector effect models (see an example in Prod'homme et al., these proceedings\cite{prodhomme20}). This allows astronomers to determine which detector effects will have the biggest impact on their science, as well as provides us with information about how well an effect needs to be characterized to be corrected for in a science exposure. As a final step, Pyxel contains a calibration mode, where laboratory data can be fed in and the parameters of the various models fit, allowing direct comparison between different detectors.

\section{Conclusion}
\label{conclusion}
The scientific detector systems of the three ELT first light instruments, HARMONI, MICADO, and METIS require a coordinated and systematic program of design, development, testing, and modeling to ensure that the varying scientific requirements can be met. The co-development and standardization of the detector systems for all three instruments will allow for streamlining of many parts of our test program while still retaining the flexibility to optimize performance for each instrument's scientific requirements. Additionally, the standardization of the test program will also ensure that any new information learned in detector testing will be immediately beneficial to all three instruments.

\acknowledgments % equivalent to \section*{ACKNOWLEDGMENTS}       
 
The authors of this paper are the people at ESO directly working on the ELT scientific detector systems. However, the work would not be possible without the support of many more of our colleagues at ESO and in the instrument consortia. We would like to acknowledge these people for their support over the years. In particular we would like to thank the team at ESO's observatories who operate and maintain the detector systems on the telescopes. We would also like to acknowledge the FIAT team, who have been working tirelessly to integrate and deliver FIAT for ELT detector testing.

% References
\bibliography{report} % bibliography data in report.bib

\begin{thebibliography}{10}

\bibitem{ramsay20}
Ramsay, S., Amico, P., Bezawada, N., Cirasuolo, M., Derie, F., Egner, S.,
  George, E., Gont\'{e}, F., Herrera, J. C.~G., Hammersley, P., Haupt, C.,
  Heijmans, J., Ives, D., Jakob, G., Kerber, F., Koehler, B., Mainieri, V.,
  Manescau, A., Oberti, S., Padovani, P., Peroux, C., Siebenmorgen, R., Tamai,
  R., and Vernet, J., ``{The ESO Extremely Large Telescope instrumentation
  programme},'' in [{\em Advances in Optical Astronomical Instrumentation
  2019}{\nolinebreak\hspace{0.1em}]},  Ellis, S.~C. and d'Orgeville, C., eds.,
  {\bf 11203},  1 -- 4, International Society for Optics and Photonics, SPIE
  (2020).

\bibitem{ramsay21}
{Ramsay}, S., {Cirasuolo}, M., {Amico}, P., {Bezawada}, N., {Caillier}, P.,
  {Deriel}, F., {Dorn}, R., {Egner}, S., {George}, E., {Gont\'{e}}, F.,
  {Hammersly}, P., {Haupt}, C., {Ives}, D., {Jakob}, G., {Kerber}, F.,
  {Mainieri}, V., {Manescau}, A., {Obertil}, S., {Peroux}, C., {Seeman}, U.,
  {Siebenmorgen}, R., {Schmidl}, C., {Vernet}, J., and {the ESO Followup team},
  ``{Instrumentation for ESO’s Extremely Large Telescope},'' {\em The
  Messenger}~{\bf 182} (Dec. 2020).

\bibitem{marchetti20}
Marchetti, E., Accardo, M., Amico, P., Brinkmann, M., Ralf~Conzelmann, M.~D.,
  Engelhardt, M., Geimer, C., Hopgood, J., Mehrgan, L., Molina, I., Mueller,
  E., Quentin, J., Reyes, J., Richerzhagen, M., Seidel, M., Stegmeier, J., and
  Todorovic, M., ``{Update on development of WFS cameras at ESO for the ELT},''
  in [{\em Adaptive Optics Systems VII}{\nolinebreak\hspace{0.1em}]},
  International Society for Optics and Photonics, SPIE (2020).

\bibitem{reyes20}
Reyes, J. and {ESO Detector Systems Group}, ``{ESO new generation wave front
  sensor cameras for the ELT: An overview and detailed description of ESO ALICE
  and LISA AO cameras},'' in [{\em Adaptive Optics Systems
  VII}{\nolinebreak\hspace{0.1em}]},  International Society for Optics and
  Photonics, SPIE (2020).

\bibitem{lucsanyi18}
{Lucsanyi}, D., {Prod'homme}, T., {Smit}, H., {Lemmel}, F., {Crouzet}, P.-E.,
  {Verhoeve}, P., and {Shortt}, B., ``{Pyxel: a novel and multi-purpose
  Python-based framework for imaging detector simulation},'' in [{\em High
  Energy, Optical, and Infrared Detectors for Astronomy
  VIII}{\nolinebreak\hspace{0.1em}]},  {\em Society of Photo-Optical
  Instrumentation Engineers (SPIE) Conference Series} {\bf 10709},  107091A
  (Jul 2018).

\bibitem{prodhomme20}
Prod'homme, T., Lemmel, F., Arko, M., Serra, B., George, E., Biancalani, E.,
  Smit, H., and Lucsanyi, D., ``{Pyxel: a novel and multi-purpose Python-based
  framework for imaging detector simulation},'' in [{\em High Energy, Optical,
  and Infrared Detectors for Astronomy IX}{\nolinebreak\hspace{0.1em}]},  {\em
  Society of Photo-Optical Instrumentation Engineers (SPIE) Conference Series}
  (2020).

\bibitem{thatte20}
{Thatte}, N.~A., {Clarke}, F., {Bryson}, I., {Shnetler}, H., {Tecza}, M.,
  {Fusco}, T., {Bacon}, R.~M., {Richard}, J., {Mediavilla}, E., {Neichel}, B.,
  {Arribas}, S., {Garcia-Lorenzo}, B., {Evans}, C.~J., {Remillieux}, A., {El
  Madi}, K., {Herreros}, J.~M., {Melotte}, D., {O'Brien}, K., {Tosh}, I.~A.,
  {Vernet}, J., {Hammersley}, P., {Ives}, D.~J., {Finger}, G., {Houghton}, R.,
  {Rigopoulou}, D., {Lynn}, J.~D., {Allen}, J.~R., {Zieleniewski}, S.~D.,
  {Kendrew}, S., {Ferraro-Wood}, V., {P{\'e}contal-Rousset}, A., {Kosmalski},
  J., {Laurent}, F., {Loupias}, M., {Piqueras}, L., {Renault}, E., {Blaizot},
  J., {Daguis{\'e}}, E., {Migniau}, J.-E., {Jarno}, A., {Born}, A., {Gallie},
  A.~M., {Montgomery}, D.~M., {Henry}, D., {Schwartz}, N., {Taylor}, W.,
  {Zins}, G., {Rodr{\'\i}guez-Ramos}, L.~F., {Cagigas}, M., {Battaglia}, G.,
  {Rebolo L{\'o}pez}, R., {Hern{\'a}ndez Su{\'a}rez}, E., {Gigante-Ripoll},
  J.~V., {Piqueras L{\'o}pez}, J., {Villa Martin}, M., {Correia}, C., {Pascal},
  S.~r., {Blanco}, L., {Vola}, P., {Epinat}, B., {Peroux}, C., {Vigan}, A.,
  {Dohlen}, K., {Sauvage}, J.-F., {Lee}, M., {Carlotti}, A., {Verinaud}, C.,
  {Morris}, T., {Myers}, R., {Reeves}, A., {Swinbank}, M., {Calcines}, A., and
  {Larrieu}, M., ``{HARMONI – first light spectroscopy for the ELT:
  quantitative performance predictions and instrument final design},'' in [{\em
  Ground-based and Airborne Instrumentation for Astronomy
  VIII}{\nolinebreak\hspace{0.1em}]},  {\em Society of Photo-Optical
  Instrumentation Engineers (SPIE) Conference Series} (2020).

\bibitem{ives20}
Ives, D., Alvarez, D., George, E.~M., Bezawada, N., Serra, B., Conzelmann, R.,
  Mueller, E., and {ESO Detector Group}, ``Characterization, operation, and
  performance aspects of h4rg-15 detectors,'' in [{\em High Energy, Optical,
  and Infrared Detectors for Astronomy IX}{\nolinebreak\hspace{0.1em}]},
  International Society for Optics and Photonics, SPIE (2020).

\bibitem{davies18}
{Davies}, R., {Alves}, J., {Cl{\'e}net}, Y., {Lang-Bardl}, F., {Nicklas}, H.,
  {Pott}, J.~U., {Ragazzoni}, R., {Tolstoy}, E., {Amico}, P.,
  {Anwand-Heerwart}, H., {Barboza}, S., {Barl}, L., {Baudoz}, P., {Bender}, R.,
  {Bezawada}, N., {Bizenberger}, P., {Boland}, W., {Bonifacio}, P., {Borgo},
  B., {Buey}, T., {Chapron}, F., {Chemla}, F., {Cohen}, M., {Czoske}, O.,
  {D{\'e}o}, V., {Disseau}, K., {Dreizler}, S., {Dupuis}, O., {Fabricius}, M.,
  {Falomo}, R., {Fedou}, P., {F{\"o}rster Schreiber}, N., {Garrel}, V., {Geis},
  N., {Gemperlein}, H., {Gendron}, E., {Genzel}, R., {Gillessen}, S.,
  {Gl{\"u}ck}, M., {Grupp}, F., {Hartl}, M., {H{\"a}user}, M., {Hess}, H.~J.,
  {Hofferbert}, R., {Hopp}, U., {H{\"o}rmann}, V., {Hubert}, Z., {Huby}, E.,
  {Huet}, J.~M., {Hutterer}, V., {Ives}, D., {Janssen}, A., {Jellema}, W.,
  {Kausch}, W., {Kerber}, F., {Kravcar}, H., {Le Ruyet}, B., {Leschinski}, K.,
  {Mandla}, C., {Manhart}, M., {Massari}, D., {Mei}, S., {Merlin}, F., {Mohr},
  L., {Monna}, A., {Muench}, N., {M{\"u}ller}, F., {Musters}, G., {Navarro},
  R., {Neumann}, U., {Neumayer}, N., {Niebsch}, J., {Plattner}, M.,
  {Przybilla}, N., {Rabien}, S., {Ramlau}, R., {Ramos}, J., {Ramsay}, S.,
  {Rhode}, P., {Richter}, A., {Richter}, J., {Rix}, H.~W., {Rodeghiero}, G.,
  {Rohloff}, R.~R., {Rosensteiner}, M., {Rousset}, G., {Schlichter}, J.,
  {Schubert}, J., {Sevin}, A., {Stuik}, R., {Sturm}, E., {Thomas}, J., {Tromp},
  N., {Verdoes-Kleijn}, G., {Vidal}, F., {Wagner}, R., {Wegner}, M.,
  {Zeilinger}, W., {Ziegleder}, J., {Ziegler}, B., and {Zins}, G., ``{The
  MICADO first light imager for the ELT: overview, operation, simulation},'' in
  [{\em Ground-based and Airborne Instrumentation for Astronomy
  VII}{\nolinebreak\hspace{0.1em}]},  {\em Society of Photo-Optical
  Instrumentation Engineers (SPIE) Conference Series} {\bf 10702},  107021S
  (Jul 2018).

\bibitem{ciliegi18}
Ciliegi, P., Diolaiti, E., Abicca, R., Agapito, G., Aliverti, M., Arcidiacono,
  C., Auricchio, N., Balestra, A., Baruffolo, A., Bellazzini, M., Bonaglia, M.,
  Bregoli, G., Brissaud, O., Busoni, L., Carlotti, A., Cascone, E., Correia,
  J.-J., Cortecchia, F., Cosentino, G., D'Orazi, V., Dall'Ora, M., Caprio,
  V.~D., Rosa, A.~D., Delboulb\'e, A., Antonio, I.~D., Rico, G.~D., Dolci, M.,
  Esposito, S., Fantinel, D., Feautrier, P., Fiorentino, G., Foppiani, I.,
  Giro, E., Gluck, L., Grani, P., Greggio, D., H\'enault, F., Jocou, L., Penna,
  P.~L., Lafrasse, S., Lauria, M., Coarer, E.~L., Louarn, M.~L., Lombini, M.,
  Magnard, Y., Magrin, D., Maiorano, E., Mannucci, F., Marchetti, E., Maurel,
  D., Michaud, L., Moraux, E., Morgante, G., Moulin, T., Oberti, S., Pariani,
  G., Patti, M., Plantet, C., Podio, L., Puglisi, A., Rabou, P., Ragazzoni, R.,
  Redaelli, E., Riva, M., Rochat, S., Roussel, F., Roux, A., Salasnich, B.,
  Saracco, P., Schreiber, L., Spavone, M., Stadler, E., Sztefek, M.-H.,
  Terenzi, L., Valentini, A., Ventura, N., V\'erinaud, C., and Zaggia, S.,
  ``{MAORY for ELT: preliminary design overview},'' in [{\em Adaptive Optics
  Systems VI}{\nolinebreak\hspace{0.1em}]},  Close, L.~M., Schreiber, L., and
  Schmidt, D., eds.,  {\bf 10703},  336 -- 345, International Society for
  Optics and Photonics, SPIE (2018).

\bibitem{marconi20}
{Marconi}, A., {Abreu}, M., {Adibekyan}, V., {Aliverti}, M., {Allende Prieto},
  C., {Amado}, P.~J., {Amate}, M., {Artigau}, E., {Augusto}, S.~R., {Barros},
  S., {Becerril}, S., {Benneke}, B., {Bergin}, E., {Berio}, P., {Bezawada}, N.,
  {Boisse}, I., {Bonfils}, X., {Bouchy}, F., {Broeg}, C., {Cabral}, A.,
  {Calvo-Ortega}, R., {Canto Martins}, B.~L., {Chazelas}, B., {Chiavassa}, A.,
  {Christensen}, L.~B., {Cirami}, R., {Coretti}, I., {Cristiani}, S., {Cunha
  Parro}, V., {Cupani}, G., {de Castro Leao}, I., {Renan de Medeiros}, J.,
  {Furlan de Souza}, M.~A., {Di Marcantonio}, P., {Di Varano}, I., {D'Odorico},
  V., {Doyon}, R., {Drass}, H., {Figueira}, P., {Fragoso}, A.~B., {Uldall
  Fynbo}, J.~P., {Gallo}, E., {Genoni}, M., {Gonzalez Hernandez}, J.~I.,
  {Haehnelt}, M., {Hlavacek Larrondo}, J., {Hughes}, I., {Huke}, P.,
  {Humphrey}, A., {Kjeldsen}, H., {Korn}, A., {Kouach}, D., {Landoni}, M.,
  {Liske}, J., {Lovis}, C., {Lunney}, D., {Maiolino}, R., {Malo}, L.,
  {Marquart}, T., {Martins}, C. J.~A.~P., {Mason}, E., {Monnier}, J.,
  {Monteiro}, M.~A., {Mordasini}, C., {Morris}, T., {Murray}, G.~J.,
  {Niedzielski}, A., {Nunes}, N., {Oliva}, E., {Origlia}, L., {Palle}, E.,
  {Pariani}, G., {Parr-Burman}, P., {Penate}, J., {Pepe}, F., {Pinna}, E.,
  {Piskunov}, N., {Rasilla}, J.~L., {Rees}, P., {Rebolo}, R., {Reiners}, A.,
  {Riva}, M., {Rousseau}, S., {Sanna}, N., {Santos}, N.~C., {Sarajlic}, M.,
  {Shen}, T.-C., {Sortino}, F., {Sosnowska}, D., {Sousa}, S., {Stempels}, E.,
  {Strassmeier}, K.~G., {Tenegi}, F., {Tozzi}, A., {Udry}, S., {Valenziano},
  L., {Vanzi}, L., {Weber}, M., {Woche}, M., {Xompero}, M., {Zackrisson}, E.,
  and {Rosa Zapatero Osorio}, M., ``{HIRES, the high-resolution spectrograph
  for the ELT},'' {\em arXiv e-prints} ,  arXiv:2011.12317 (Nov. 2020).

\bibitem{jagourel18}
{Jagourel}, P., {Fitzsimons}, E., {Hammer}, F., {De Frondat}, F., {Puech}, M.,
  {Evans}, C.~J., {Sanchez}, R., {Guinouard}, I., {Chemla}, F., {Frotin}, M.,
  {Yang}, Y., {Parr-Burman}, P., {Morris}, T., {Dubbeldam}, M., {Close}, M.,
  {Middleton}, K., {Rousset}, G., {Gendron}, {\'E}., {Kelz}, A., {Janssen}, A.,
  {Pragt}, J., {Navarro}, R., {Larrieu}, M., {El Hadi}, K., {Dohlen}, K.,
  {Dalton}, G., {Lewis}, I., {Rodrigues}, M., {Morris}, S., {Kaper}, L.,
  {Barbuy}, B., {Cuby}, J.~G., and {Le F{\`e}vre}, O., ``{MOSAIC: the ELT
  multi-object spectrograph},'' in [{\em Ground-based and Airborne
  Instrumentation for Astronomy VII}{\nolinebreak\hspace{0.1em}]},  {Evans},
  C.~J., {Simard}, L., and {Takami}, H., eds., {\em Society of Photo-Optical
  Instrumentation Engineers (SPIE) Conference Series} {\bf 10702},  10702A4
  (July 2018).

\bibitem{hammer20}
{Hammer}, F., {Morris}, S., {Cuby}, J.~G., {Kaper}, L., {Steinmetz}, M.,
  {Afonso}, J., {Barbuy}, B., {Bergin}, E., {Finogenov}, A., {Gallego}, J.,
  {Kassin}, S., {Miller}, C., {Ostlin}, G., {Penterricci}, L., {Schaerer}, D.,
  {Ziegler}, B., {Chemla}, F., {Dalton}, G., {De Frondat}, F., {Evans}, C., {Le
  Mignant}, D., {Puech}, M., {Rodrigues}, M., {Sanchez-Janssen}, R., {Taburet},
  S., {Tasca}, L., {Yang}, Y.~B., {Zanchetta}, S., {Dohlen}, K., {Dubbeldam},
  M., {El Hadi}, K., {Janssen}, A., {Kelz}, A., {Larrieu}, M., {Lewis}, I.,
  {MacIntosh}, M., {Morris}, T., {Navarro}, R., and {Seifert}, W., ``{MOSAIC on
  the ELT: high-multiplex spectroscopy to unravel the physics of stars and
  galaxies from the dark ages to the present-day},'' {\em arXiv e-prints} ,
  arXiv:2011.03549 (Nov. 2020).

\bibitem{bezawada20}
Bezawada, N., Ives, D., Alvarez, D., Serra, B., George, E., Mandla, C., and
  Mehrgan, L., ``Performance advantages of buffered mode operation of hxrg near
  infrared detectors,'' in [{\em High Energy, Optical, and Infrared Detectors
  for Astronomy IX}{\nolinebreak\hspace{0.1em}]},  International Society for
  Optics and Photonics, SPIE (2020).

\bibitem{lizon16}
Lizon, J.~L., Amico, P., Brinkmann, M., Delabre, B., Finger, G., Guidolin,
  I.~M., Guzman, R., Hinterschuster, R., Ives, D., Klein, B., and Quattri, M.,
  ``{A new test facility for the E-ELT infrared detector program},'' in [{\em
  Ground-based and Airborne Instrumentation for Astronomy
  VI}{\nolinebreak\hspace{0.1em}]},  Evans, C.~J., Simard, L., and Takami, H.,
  eds.,  {\bf 9908},  2857 -- 2866, International Society for Optics and
  Photonics, SPIE (2016).

\end{thebibliography}
\bibliographystyle{spiebib} % makes bibtex use spiebib.bst

\end{document}